\documentclass[debug]{rmaa}
\usepackage{natbib}

\usepackage[latin1]{inputenc}

\title{Entrainment factor on individual glitch fractional moment of inertia} 

\author{
  I. O. Eya,\altaffilmark{1,3} 
 J. O. Urama,\altaffilmark{2,3}
  and A. E. Chukwude\altaffilmark{1,2,3}}

\altaffiltext{1}{Department of Science Laboratory Technology, University of Nigeria, Nsukka, Nigeria.}

\altaffiltext{2}{Department of Physics \& Astronomy, University of Nigeria, Nsukka, Nigeria.}

\altaffiltext{3}{Astronomy and Astrophysics Research Center, Faculty of Physical Sciences, University of Nigeria, Nsukka, Nigeria.}

\shortauthor{Eya, Urama \& Chukwude}
\shorttitle{Entrainment factor on individual glitch size}

\fulladdresses{
\item Eya, I. O: Department of Science Laboratory Technology, University of Nigeria, Nsukka, Nigeria. (innocent.eya@unn.edu.ng).
\item Urama, J. O: Department of Physics and Astronomy, University of Nigeria, Nsukka, Nigeria. (johnson.urama@unn.edu.ng)
\item Chukwude, A. E: Department of Physics and Astronomy, University of Nigeria, Nsukka, Nigeria.
  (augustine.chukwude@unn.edu.ng)}

\listofauthors{Eya, I. O. Urama, J. O. \& Chukwude, A. E.}
\indexauthor{Eya, I. O.}
\indexauthor{Urama, J. O.}
\indexauthor{Chukwude, A. E.}

\abstract{
The superfluid in the inner crust of neutron star is assumed to be the reservoir of momentum
released in pulsar glitch. Recently, due to crustal entrainment, it is debatable whether the
magnitude of the inner crust is sufficient to contain superfluid 
responsible for large glitches. This paper calculates the fractional moment of inertia
(FMI)(i.e. the ratio of the inner crust superfluid moment of inertia to that of the coupled components)
associated with individual glitches. 
It is shown that the effective moment of inertia associated with the transferred momentum is that of the entrained neutrons.  
The FMI for glitches in three pulsars, which exhibit the signature of exhausting their momentum reservoir
were calculated and scaled with entrainment factor. 
Some of the glitches require inner crust superfluid with moment of
inertia larger than the current suggested values of 7-10\% of the stellar moment of inertia.}

\addkeyword{methods: statistical}
\addkeyword{pulsars: general}
\addkeyword{stars: neutron}

\begin{document}
\maketitle

\section{Introduction}
\label{sec:intro}

Pulsars are spinning magnetized neutron star \citep{b8a}. 
The spin rate of these objects are highly stable due to huge moment of inertia they posses ($ \approx 10^{45}$ g cm$^{2} $). 
In spite of this, the spin rate of some pulsars is occasionally perturbed in events known as glitch. 
Pulsar glitches are impulsive increase in pulsar spin frequency, $ \Delta \nu $ \citep{b12,b31,b7,b33}.
This event is sometimes associated with change in pulsar spin-down rate, $ \Delta \dot{\nu} $ \citep{b9a}. 
In most of the pulsars, glitch events are believed to involve superfluid neutrons in the inner crust of neutron star \citep{b5,b3,b1}.
Firstly, this is due to the long time it takes a pulsar to recover to a steady spin frequency after a glitch (days to months) and, secondly, due to recovery phase, which is exponential in nature for most pulsars.
Recently, interior of neutron star containing superfluid have gained observational evidence in cooling of young neutron stars \citep{b10,b14}.
So it is not in doubt if neutron star contains a superfluid component. 
 
Pulsar glitch models involving superfluid component view the neutron star as a system in which its components rotate differentially. 
The main components are: the solid crust, the interior superfluid neutrons (inner crust and outer core), and the core \citep{b15}.
In this model, the solid crust and the core are coupled electromagnetically. 
The inner crust superfluid component viewed as momentum reservoir, rotates via array of quantized vortices whose areal density is proportional to the fluid velocity. 
These vortices are pinned in the ion lattice of the inner crust, leading to partial decoupling of the inner crust superfluid component from the other components \citep{b3,b1}.
As the coupled components spins-down electromagnetically, the inner crust superfluid maintains its own velocity.
In this situation, the superfluid at a higher velocity stores angular momentum, which is occasionally released in glitches.
For the superfluid to spin-down, the vortex areal density must decrease.
This could be either by reduction in vortex number, or by outward migration of vortices.
Such processes is prevented by the pinning force on the vortices.
As long as the vortices remain in their pinned position, the superfluid angular momentum is conserved.

Meanwhile, as the solid crust lags behind the superfluid component, the rotation lag (i.e. the magnitude of the velocity difference between the two components) increases with time.
The lag is not sustainable for pulsar life time. 
At a critical lag, unclear mechanism unpins some of the vortices (or unpinning of the entire vortices). 
The vortices migrate outward transferring their momentum to the crust; the superfluid spins-down and the crust spins-up \citep{b3,b1}.
The magnitude of the crustal spin-up, $ \Delta \nu $, is the glitch spin-up size.
Large glitches such as that of Vela pulsar are characterised by $ \Delta \nu > 10^{-6} \ Hz $. 
Such a glitch size is one of the reasons why scholars believed that there is angular momentum reservoir somewhere in neutron star interior.

Glitch model involving angular momentum transfer is standard for discussing pulsar glitches for decades.
This is partly due to its ability to explain post glitch features such as exponential recoveries and long recovery times \citep{b5,b1}, and mainly due to agreement between theoretical prediction of neutron star crustal thickness and pulsar glitch size 
\citep{b13,b8}.
Recently, most aspect of Vela pulsar glitches have been fully described based on this model \citep{b8b}, that plausibly, angular momentum transfer model is attaining a status of standalone theory.
In the work of \citet{b8}, the moment of inertia of the superfluid component involved in Vela glitches is just about 1.4\% of the stellar moment of inertia. 
This amount of superfluid can conveniently reside in the inner crust of the star.
In view of the inner crust superfluid involvement in pulsar glitches, the regularity of glitches in Vela pulsar and PSR J$0357-6910$, is seen as a consequence of recycling a reservoir that is exhausted at each event \citep{b4}.

However, following recent involvement of crustal entrainment in pulsar glitch size \citep{b4,b6}, angular momentum transfer models is under a serious challenge.
Basically entrainment increases the inertia of superfluid neutrons, thereby reducing the mobility of the fluid \citep{b6a,b6d,b6b}.
For this reason, the superfluid confined in the inner crust is not sufficient to produce Vela-like glitches; unless glitching pulsars are low mass neutron stars ($ \leq 1.0 \ M_{\odot} $), or that the core fluid is involved in the glitch \citep{b4}.
Consequently, in \citet{b8} the inner crust superfluid moment of inertia is underestimated by a factor of 4.3 \citep{b4,b6}, which is the likely value of entrainment factor.
Physically, it means that the moment of inertia of the superfluid contains in the inner crust should be above $ 6\% $ of the stellar moment of inertia for it to produce the observed glitches (i.e. $4.3\times 1.4\% $).

On the other hand, recent works \citep{b11,b13a} have argued that the inner crust superfluid could sufficiently produce the observed glitches. 
The argument is based on exploring the uncertainties in nuclear matter Equation-of-States (EoS), which models the structure of neutron star. 
With this approach, \citet{b11} obtain a crust that is thick enough to contain fluid of up to 7\% stellar moment of inertia given neutron star mass of $ < 1.6\ M_{\odot} $. 
Similarly, for neutron star mass of $ 1.4\ M_{\odot} $, \citet{b13a} obtained a thicker crust that is up to 10\% stellar moment of inertia. 
Large crustal thickness implies large stellar radius and small stellar mass. 
In this frame, there is a limit one can extend the crust no matter the magnitude of uncertainty in the EoS, else we will be tilting towards white dwarf. 

In the previous analyses, the approach has been calculating the fractional moment of inertia (FMI) (i.e. the ratio of inner crust superfluid moment of inertia to that of the coupled components) of neutron star components participating in glitch based on average glitch size in a given pulsar.
The result is then compared with the theoretical magnitude of neutron star crustal thickness.  
Inasmuch as this approach is fair, it hides the intrinsic size of the inner crust fluid. 
Effort should be channelled towards calculating the FMI based on individual glitches, as this will show the possible range of crustal thickness.  
In this analysis, this paper treats each glitch as a unique event, and calculated the FMI for each glitch in three pulsars that exhibit strong linear transfer of angular momentum with time.
The linearity of glitches in these pulsars is believed to be a consequence of a reservoir that is exhausted at each event, thereby making each glitch independent of one another.
In such a situation, the FMI for each glitch is a measure of distinct momentum reservoir.
The result shows that some glitches exceed the initial inner crust moment of inertia as constrained in \citet{b8} even without the entrainment factor. 
In addition, if the entrainment factor stands at 4.3, the present neutron star crustal thickness ($\approx 10\%$) is not sufficient to produce some glitches.  

\section{Rotation lag and fractional moment of inertia}
For a spinning neutron star, the standard rotation lag between the momentum reservoir\footnote{the superfluid confined in the inner crust} and the observable solid crust, which leads to accumulation of transferable momentum is
\begin{equation}
\omega(t) = \Omega_{res} - \Omega_{c}(t),
\end{equation}
where $ \Omega_{res} = 2\pi\nu_{res} $ is the reservoir's angular frequency,
$ \Omega_{c} = 2\pi\nu_{c} $ the angular frequency of the solid crust and any other component coupled to it.
The stellar moment of inertia is
\begin{equation}
I = I_{res} + I_{c},
\end{equation}
where $ I_{res} $ and $ I_{c} $ are the moments of inertia of the momentum reservoir and that of the coupled components respectively.
In this model, $ I_{c} $ make up at least 90\% of the neutron star moment of inertia \citep[and references therein]{b8}, implying $ I_{c} \approx I $.

In a glitching pulsar, at a time-interval ($ t_{i} $) preceding a glitch, the reservoir accumulates transferable momentum due to the rotation lag, which can be quantified by 
\begin{equation}
L_{i} = I_{res}\omega(t),
\end{equation} 
at a rate of
\begin{equation}
\dot{L}_{i} = -I_{res}\dot{\Omega}_{ic}(t),
\end{equation}
where $ \dot{\Omega}_{ic}(t) = 2\pi\dot{\nu}_{ic}$ is the spin-down rate of the crust at a time-interval preceding the glitch.
Here, it is assumed that the momentum accumulated over a period $ t_{i} $, manifests in spin-up of the crust $ \Delta\Omega_{ic} $. 
A measure of $ \Delta\Omega_{c} $ is an indirect way of estimating the transferred momentum.
In this, for a given glitch, the transferred momentum is
\begin{equation}
L_{i} = I_{c}\Delta\Omega_{ic}(t),
\end{equation}
at a rate of
\begin{equation}
\dot{L}_{i} = I_{c}\frac{\Delta\Omega_{ic}(t)}{t_{i}}.
\end{equation}

In this frame, if the rate of accumulation of angular momentum by the reservoir is directly proportional to the rate angular momentum is transferred, the cumulative glitch spin-up sizes ($ \Sigma\Delta\Omega_{ic} $) should be linear with time if the momentum reservoir is exhausted at each glitch.  
Such pulsars of linear transfer of angular momentum with time are shown in Fig. 1.
This kind of behaviour has been reported in PSRs J$0835-4510 $ (Vela pulsar) \citep{b8,b9,beu},  J$0537-6910 $ \citep{b43a}, and J$1420-6048 $ \citep{b7e}.
Hence, Equation (4) and (6) gives the individual glitch FMI
\begin{equation}
\frac{I_{res}}{I_{c}} = - \frac{1}{{\dot{\Omega}}_{ic}(t)}\frac{\Delta\Omega_{ic}}{t_{i}}(t).
\end{equation}
Such an expression for FMI has been obtained earlier \citep{b7e}.
The magnitude of FMI gives an insight on the magnitude of the momentum reservoir.

\section{Entrainment factor and the magnitude of FMI/glitch size}
It is known that superfluid flows with zero viscosity.
The superfluid neutrons in the inner crust of a neutron star also flows with zero viscosity, but it is still entrained by the crust \citep{b10a}.
The entrainment is non-dissipative, it occurs due to the elastic scattering of free neutrons by crustal lattice \citep{b6}.
The magnitude of entrainment in the inner crust is quantified by either the density of conduction neutrons in the crust or by the effective mass of neutron \citep{b4,b6}.
In this paper, the interest is on how the entrainment factor constrains the observed glitch sizes.

For a sphere spinning down, such as pulsars, the loss in rotational energy is
\begin{equation}
\dot{E}= I\Omega_{c}\dot{\Omega}_{c}.
\end{equation}
This loss in rotational energy manifest in the observed radiation from the pulsar, which can be approximated to that of a dipole radiator in a vacuum,
\begin{equation}
\dot{E}= - \frac{B^{2}R^{6}\Omega^{4}}{6c^{3}}\sin\alpha^{2},
\end{equation}
where $B $ is the magnetic field strength, $ R $ is the stellar radius, and $ c$ is the speed of light.
Comparing Equations (8) and (9) (with $ \Omega\equiv\Omega_{c} $) lead to
\begin{equation}
I\dot{\Omega}_{c} = - K\Omega^{3}_{c}.
\end{equation}
Equation (10) is the standard spin-down law of pulsars, where $ K =  6^{-1}B^{2}R^{6}$ $c^{-3} \sin^{2} \alpha $ is assumed to be constant.
For the two components model and owing to the pinned vortices, 
\begin{equation}
I_{c}\dot{\Omega}_{c} + I_{res}\dot{\Omega}_{res}= - K\Omega^{3}_{c}.
\end{equation}
As entrainment is non-dissipative, $ I_{res}\dot{\Omega}_{res} $ is not expected to affect the spin down of the pulsar.
In the frame of perfect pinning and expressing entrainment in terms of coefficient, $ e_{n} $, \citep[and reference therein]{b4},
\begin{equation}
I_{res}\dot{\Omega}_{res} = -\frac{e_{n} I_{res}\dot\Omega_{c}}{(1-e_{n})}.
\end{equation}
Therefore, the effective torque on the pulsar is 
\begin{equation}
I_{ef}\dot{\Omega}_{c} = - K\Omega^{3},
\end{equation}
where $ I_{ef}=I_{c}-(\frac{e_{n}}{1-e_{n}})I_{res} $ is the effective moment of inertia of the pulsar due to entrainment, and  $ (\frac{e_{n}}{1-e_{n}}) = E_{n}$ is the entrainment factor.
If the entrainment coefficient is zero, the standard spin-down law is recovered. 

Let us determine effective moment of inertia associated with the glitch event.
Based on the two component model, the total angular momentum of the system is
\begin{equation}
L_{tot} = I_{res}\Omega_{res} + I_{c}\Omega_{c},
\end{equation}
where $ I_{res}\Omega_{res} $ is the angular momentum of the momentum reservoir (neutron superfluid), and $ I_{c}\Omega_{c} $ is the observable angular momentum of the star.
However, owing to entrainment, the angular momentum of the superfluid is a function of both the superfluid angular velocity and the angular velocity of the star, $ \Omega_{c} $, \citep{b6}, which is expressed as \citep{b6d,b6}
\begin{equation}
L = I_{ss}\Omega_{res} + (I_{res}-I_{ss})\Omega_{c},
\end{equation}
where $ I_{ss} $ is the moment of inertia associated with the entrained neutrons.
Hence, the total angular momentum of the system as a result of entrainment is\footnote{we have made use of $ I = I_{c} + I_{res} $}
\begin{equation}
L_{tot} = I_{ss}(\Omega_{res}-\Omega_{c}) + I\Omega_{c}.
\end{equation}
The term in parenthesis is the differential rotational lag, $\omega(t) $.
The implication of this is that the effective moment of inertia associated with the transferred momentum is that of the entrained neutron, and $ I_{res} $ in Equation (7) could be safely replaced with $ I_{ss} $.
From Equation (16), the effective torque is $ (I-I_{ss})\dot{\Omega}_{c} $ and with Equation (13)\footnote{we have made use of $ I \approx I_{c} $} $ I_{ss} = E_{n}I_{res} $.
Hence incorporating  entrainment factor in the expression of FMI leads to, 
\begin{equation}
\frac{I_{res}}{I_{c}} =  - \frac{1}{E_{n}}\frac{1}{{\dot{\Omega}}_{ic}(t)}\frac{\Delta\Omega_{ic}}{t_{i}}(t),
\end{equation}
The interpretation of this result is quite simple; the observed glitch sizes should be lesser by a factor of $ 1/E_{n} $, or equivalently, the moment of inertia of the momentum reservoir should be enhanced by a factor of $ E_{n} $ for the observed glitch sizes.
Though this argument is not new, it has not been extended to individual glitch FMI.

\section{Data and result}
The glitches for this analysis are from \citet{b7}, and updated with JBO glitch tables and references therein\footnote{http://www.jb.man.ac.uk/pulsar/glitches.html, accessed on may 1, 2017.} to include more recent events as of time of this analysis.  
Three pulsars in which their glitch spin-up sizes ($ \Delta \nu$) are quite regular with time were selected for this analysis (Fig.1).
Concentrating on such pulsars is a precaution to avoid glitches that may originate from any other component other than the crustal superfluid, which is the bases for regularity of glitches. 
In addition, pulsars in which their glitches follows this trend are believed to possess a reservoir that is exhausted at each glitch. 
In this frame, each of the glitches is a unique event independent of one another.

For nearly two decades, constraint on crustal thickness estimated from glitch data is based on comparing the moment of inertia of the inner crust superfluid obtain from a linear fit to glitch points in plots of such as Fig. 1. 
With the slope of the linear fits in Fig. 1, and the mean spin down rate of the corresponding pulsar, the mean FMI in each of the pulsars are: 0.85\% for PSR J$0537-6910 $, 1.55\% for PSR J$0835-4510$ and 1.29\% for PSR J$1420-6048 $.
These values are in line with other analyses \citep{b8,b4,b7e}.
A constraint of this kind is based on average glitch sizes in a pulsar.
This approach do not give room for the extreme values.

The FMIs for each glitch are shown in Table 1.
The FMI corresponding to the first glitch in a given pulsar could not be calculated as the time interval preceding the glitch is not available.  
The distribution of FMIs is shown in top panel of Fig. 2, while the 
bottom panel is the distribution of FMIs scaled with the entrainment factor (4.3).
From the top panel, it is clear that $ 1.4 \% $ crustal thickness moment of inertia could not accommodate the observed glitches even without entrainment factor. 
The relevant glitches are the glitches with FMIs at the right-hand-side of $ 1.4 \% $ line.
These glitches make up $\approx 52\%$ of the glitches in these pulsars.
Interestingly, when the FMIs is scaled by the entrainment factor,  
$\approx 93\%$ of the glitches requires crustal thickness that is more than $ 1.4\% $ of the stellar moment of inertia as seen in bottom panel of Fig. 2.
In addition, with entrainment factor $ \approx 26\% $ of the glitches requires a crust beyond the possible 10\% stellar moment of inertia proposed by \citet{b13a}.

\begin{figure}
\includegraphics[scale=0.8]{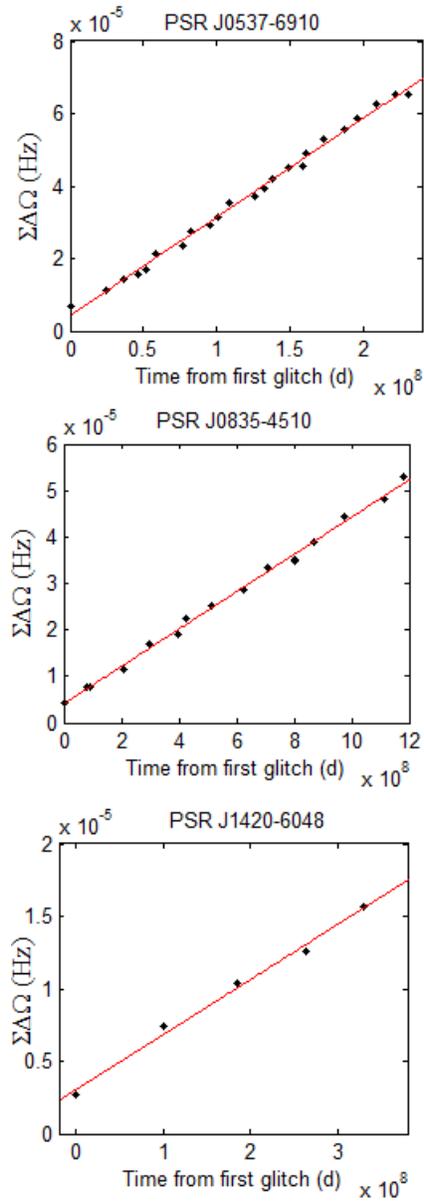}
\centering
\caption{\fontsize{8pt}{1.6}\selectfont
Regularity of pulsar glitches. The straight line is a linear fit to the points}
\end{figure}

\begin{table}
\centering
\caption{\fontsize{8pt}{1.6}\selectfont
Characteristic FMI in the pulsars.}\fontsize{8pt}{10}\selectfont
   \begin{center}  
  \begin{tabular}{@{}|cccc|@{}}
  \hline
         $ N_{g} $    &J0537-6910 & J0835-4510 & J1420-6048 \\

		\hline
	1&	---  &---&	---  \\
	2&	0.763& 1.851&  1.660\\
	3&	1.067&	0.080&	1.241\\
	4&	0.724&	1.448&	0.959\\
	5&	0.964&	2.605&	1.685\\
	6&	2.182&	0.792&	---\\
	7&	0.595&  5.570&	---\\
	8&	21.267&	1.244&	---\\
	9&	0.418&	1.184&	---\\
	10&	1.156&	2.395&	---\\
	11&	1.449&	0.627&	---\\
	12&	0.301&	5.116&	---\\
	13&	1.165&	2.241&	---\\
	14&	1.234&	2.145&	---\\
	15&	1.025&	1.093&	---\\
	16&	0.057&	2.812&	---\\
	17&	6.804&	1.095&	---\\
	18&	0.944& 2.225&	---\\
	19&	0.612& 0.001&	---\\
	20&	1.089&---&	---\\
	21&	0.990&---&	---\\
	22&	0.492&---&	---\\
	23&	0.067&---&	---\\

\hline 
\end{tabular}
\end{center}
Note: The FMIs are measured in percent (\%), $ N_{g} $ denotes the glitch number. 
\end{table}

\begin{figure}
\centering
\includegraphics[scale=0.7]{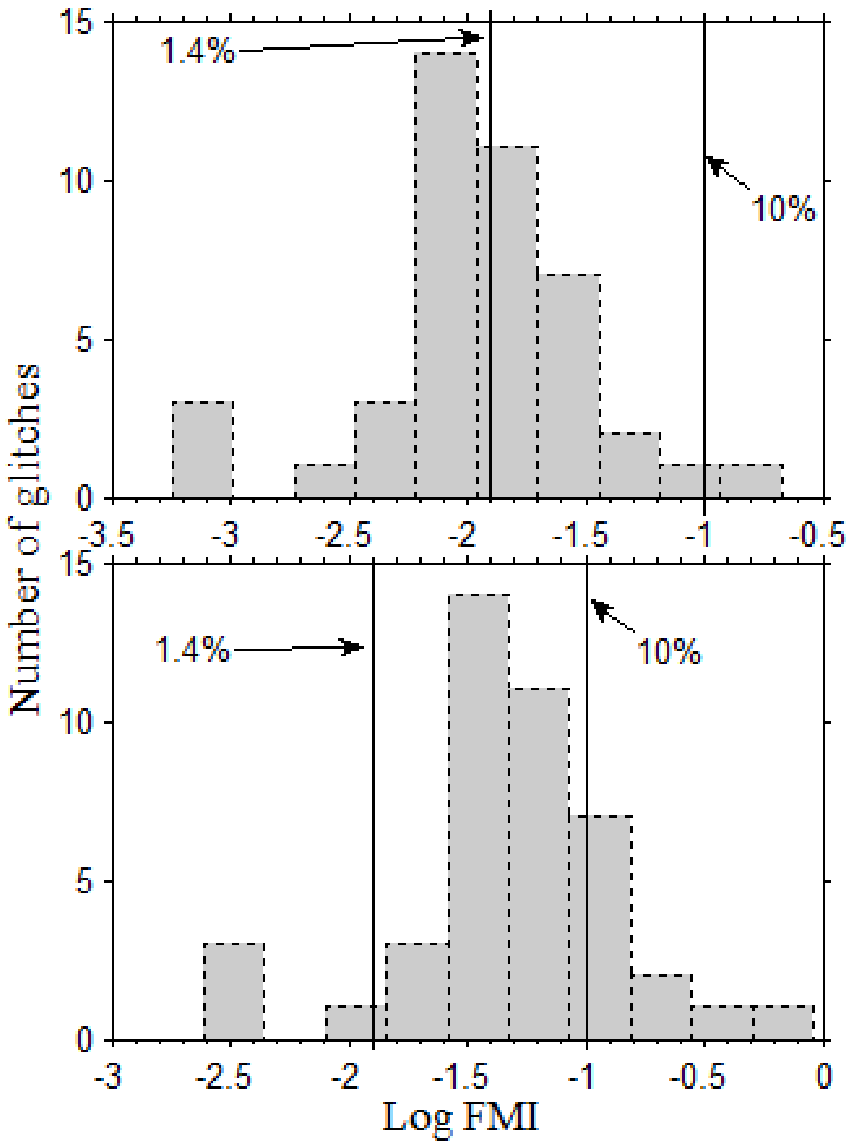}
\caption{\fontsize{8pt}{1.6}\selectfont
Distribution of FMIs calculated \textbf{from Equation (7)}; bottom panel is the distribution of FMIs scaled with entrainment factor \textbf{as suggested by Equation (17)}.}
\end{figure}

\section{DISCUSSION}
\label{sec:DISCUSSION}
The FMIs in this analysis is a measure of distinct reservoir moment of inertia.
The upper limit in the range of FMIs in a given pulsar gives an insight on the minimum size of the neutron star crustal thickness. 
Without the entrainment factor, the: $ 6th$, $8th$ and $ 17th $
glitch in PSR J$0537-6910$; $ 5th $, $7th$, $ 10th $, $ 12th $, $ 13th $, $ 14th $, $ 16th $ and $ 18th $ glitch in Vela pulsar requires a crustal thickness that is above 2\%, stellar moment of inertia. 
It is only the glitches in PSR J$1420-6048$ that are exempted from this anomaly. 
Owing to this, the earlier theoretical calculations of neutron star crust \citep{b13,b8} could not account for some of the glitches even without entrainment factor.
This effect is more severe with entrainment factor where some of the glitches will require crust thickness that is above 10\% stellar moment of inertia.
FMI could be as large as 90\% and 24\% for the $8th$ and $7th$ glitches in PSRs J$0537-6910$ and J$0835-4510$ respectively\footnote{i.e. multiplying the FMI by entrainment factor (4.3).}. 
As at present, no EoS has a neutron star crust that could contain a crustal fluid for such a reservoir. 
This result is quite disturbing if one recalls that these glitches are from pulsars, which empties their reservoir at each glitch.
There is no evidence of radiative change in these pulsars during glitch, which might have suggest that their glitches are enhanced by magnetospheric activity.
The 7-10\% crustal thickness moment of inertia \citep{b11,b13a} is an upper limit in the current theoretical calculation of neutron star structure. 
The actual value could be lower as the authors neglected superfluidity.
In particular for such a crustal thickness, the neutron star radius should be as large $\approx 14.0 \pm 0.5 $ km.
Clearly this value is in contrast with recent analysis of low mass X -ray binaries, which predicted small radii of $\approx 11.8 \pm 0.9 $ km \citep{bb8}, and even smaller from the analysis of \citet{ba8}.

Finally, pulsar glitch models relying on inner crust superfluid and nuclear matter EoS is under serious challenge unless the vortex unpinning trigger mechanism, which is still elusive has the ability to squeeze angular momentum and liberate it at the onset of glitch.

\end{document}